\documentclass[12pt]{iopart}
\usepackage{url}

\usepackage{graphicx}
\usepackage{siunitx}
\usepackage{xcolor}
\usepackage[bookmarks=false]{hyperref}
\usepackage{lineno}
\def\equationautorefname~#1\null{(#1)\null}
\def\figureautorefname~#1\null{figure~#1\null}

\begin{document}

\title{Gas cooling of test masses for future gravitational-wave observatories}

\author{Christoph Reinhardt$^1$, Alexander Franke$^2$, Jörn Schaffran$^1$, Roman Schnabel$^2$ and Axel Lindner$^1$}

\address{$^1$ Deutsches Elektronen Synchrotron (DESY), 22607 Hamburg, Germany}

\address{$^2$ Institut für Laserphysik und Zentrum für Optische Quantentechnologien der Universität Hamburg, Hamburg, Germany}

\ead{christoph.reinhardt@desy.de}
\vspace{10pt}
\begin{indented}
    \item[]June 2021
		\item[]
		\item[]\textcolor{gray}{\textit{All figures and pictures by the authors under a CC BY 4.0 license}}
\end{indented}

\begin{abstract}
    Recent observations made with Advanced LIGO and Advanced Virgo have initiated the era of gravitational-wave astronomy.
    The number of events detected by these ``2\textsuperscript{nd} Generation'' (2G) ground-based observatories is partially limited by noise arising from temperature-induced position fluctuations of the test mass mirror surfaces used for probing spacetime dynamics.
    The design of next-generation gravitational-wave observatories addresses this limitation by using cryogenically cooled test masses;
    current approaches for continuously removing heat (resulting from absorbed laser light) rely on heat extraction via black-body radiation or conduction through suspension fibres.
    As a complementing approach for extracting heat during observational runs, we investigate cooling via helium gas impinging on the test mass in free molecular flow.
    We establish a relation between cooling power and corresponding displacement noise, based on analytical models, which we compare to numerical simulations.
    Applying this theoretical framework with regard to the conceptual design of the Einstein Telescope (ET), we find a cooling power of \SI{10}{mW} at \SI{18}{K} for a gas pressure that exceeds the ET design strain noise goal by at most a factor of $\sim3$ in the signal frequency band from 3 to \SI{11}{Hz}.
    A cooling power of \SI{100}{mW} at \SI{18}{K} corresponds to a gas pressure that exceeds the ET design strain noise goal by at most a factor of $\sim11$ in the band from 1 to \SI{28}{Hz}.
\end{abstract}

%
%
%
%
%

\section{Introduction}
\label{sec:introduction}

The first detection of gravitational waves by LIGO in September 2015 has unlocked a new source of information about the universe~\cite{abbott2016observation}.
So far, LIGO together with Virgo has observed 15 confirmed events and 35 candidate events of gravitational waves originating from mergers of two black holes, two neutron stars, as well as pairs of one black hole and one neutron star~\cite{abbott2019gwtc1,abbott2020gwtc2}.
In order to increase the rate and range of detections, a ``3\textsuperscript{rd}~Generation'' (3G) of ground-based observatories is currently being developed~\cite{abernathy2011einstein,ET2020einstein,adhikari2020cryogenic,reitze2019cosmic}.
Research targets increasing the GW signal as well as reducing the observatory's detection noise floor.
The signal increases with the interferometer arm length and with the light power in the arms.
Noise sources that are going to be reduced have many origins.
The largest fundamental noise sources in the centre of the detector bandwidth ($\sim 100$\,Hz) are the quantum uncertainty in the measurement of the laser light and the thermally excited motions of the mirror surfaces.
The latter are produced by thermal energy in all the different degrees of freedom of massive test mass mirrors that are suspended as pendulums under vacuum conditions.
The most prominent example of thermal noise results from the Brownian motion within the dielectric high-reflectivity coatings of the mirrors.

Thermal noise is reduced if the temperature of the suspended test mass mirrors is lowered.
Current LIGO~\cite{aasi2015advanced} and Virgo~\cite{acernese2014advanced} observatories exploit mirrors at room temperature.
The Japanese KAGRA~\cite{somiya2012detector} observatory, which began initial observations in Feb 2020~\cite{akutsu2020overview}, exploits mirrors cooled to about \SI{20}{K}.
The designs of the European Einstein Telescope as well as LIGO Voyager and Cosmic Explorer, the U.S.\ contribution to a future 3G detector network, incorporate cryo-cooling as well.
In the range from \SI{40}{Hz} to \SI{100}{Hz}, where thermal noise is a dominating source of noise~\cite{buikema2020sensitivity,acernese2020advanced}, a significant sensitivity improvement is expected.

Cryogenic cooling of up to $\sim300~\mathrm{kg}$ mirrors that are suspended with rather thin fibres~\cite{abernathy2011einstein, ET2020einstein, reitze2019cosmic} is a major technological challenge.
The problem is how to continuously get thermal energy out of mirrors that are in vacuum and mechanically maximally decoupled from the environment, while the observatory is taking data.
Heat load due to absorbed black-body radiation from the (room-temperature) kilometre-scale vacuum tubes has to be suppressed to a minimum.
During operation, the test masses are constantly heated by partial absorption of laser light.
Mirror substrate and coating materials need to show extremely low optical absorption in the range of a few parts per million (ppm).
At the same time, the materials need to have high mechanical quality factors to channel the remaining thermal energy in narrow well-defined mechanical resonances.

In KAGRA, heat is extracted from the mirrors via suspension fibres with high thermal conductivity~\cite{akutsu2019first, ushiba2021cryogenic}.
The cooling power is provided by cryocoolers, which are connected to a separate stage of the suspension chain with flexible high-purity aluminum heat links;
a dedicated vibration isolation system has been implemented to suppress the impact from cryocooler induced vibrations.
For the cryogenic detector of ET, with an intended operating temperature in the range of \SI{10}{K} to \SI{20}{K}, a similar cooling strategy is considered~\cite{ET2020einstein}.
Cosmic Explorer, by virtue of its higher operating temperature of \SI{123}{K}, will be cooled by radiative heat transfer from the mirror to surrounding radiation shields~\cite{reitze2019cosmic}.
This approach is currently being developed in the scope of LIGO Voyager~\cite{adhikari2020cryogenic}.

Heat-exchange gas is often used for cooling down sensitive probes, which is also under investigation for GW detectors~\cite{shapiro2017cryogenically, bonilla2019improving}.
Using heat-exchange gas \emph{during} GW observations is a less obvious approach since the gas causes friction and transfers momentum to the mirror.
Previously, the 1.5 tonnes bar detector Niobe used gas cooling during observation runs~\cite{tobar1995university}.
Based on the reported parameters, we estimate the gas pressure to be above the free molecular flow regime.

Here, we investigate the potential of using helium gas in the free molecular flow regime to extract the heat imparted during observational runs in cryogenic test masses of future gravitational wave observatories.
Heat transfer via conduction in suspension fibres is not included.
Section~\ref{sec:suspended-mirror-setup} describes the conceptual setup.
It further includes a discussion of potential challenges related to additional noise sources, arising as a consequence of injecting a significant amount of helium gas next to the mirror.
Section~\ref{sec:models-for-heat-transfer-and-strain-noise} establishes, to our knowledge for the first time, the relation between gas-induced cooling power and added strain-normalized noise spectral density, based on corresponding analytical models, which are validated by comparison to numerical simulations.
Our analytical models for cooling power and residual gas damping noise are also in agreement with previous publications~\cite{corruccini1959gaseous, cavalleri2010gas}.
Consequences from gradually ``turning on'' mutual collisions between helium atoms, thereby ``leaving'' the free molecular regime, are also discussed.
Section~\ref{sec:cooling-power-and-added-thermal-noise-for-et-design} presents an application of the theoretical framework described in the previous section with regard to the design of the Einstein Telescope.
Section~\ref{sec:conclusion} gives concluding remarks.

\section{Conceptual setup for gas cooling applied to suspended test masses}
\label{sec:suspended-mirror-setup}
\begin{figure}
    \centering
    \includegraphics[width=.45\textwidth]{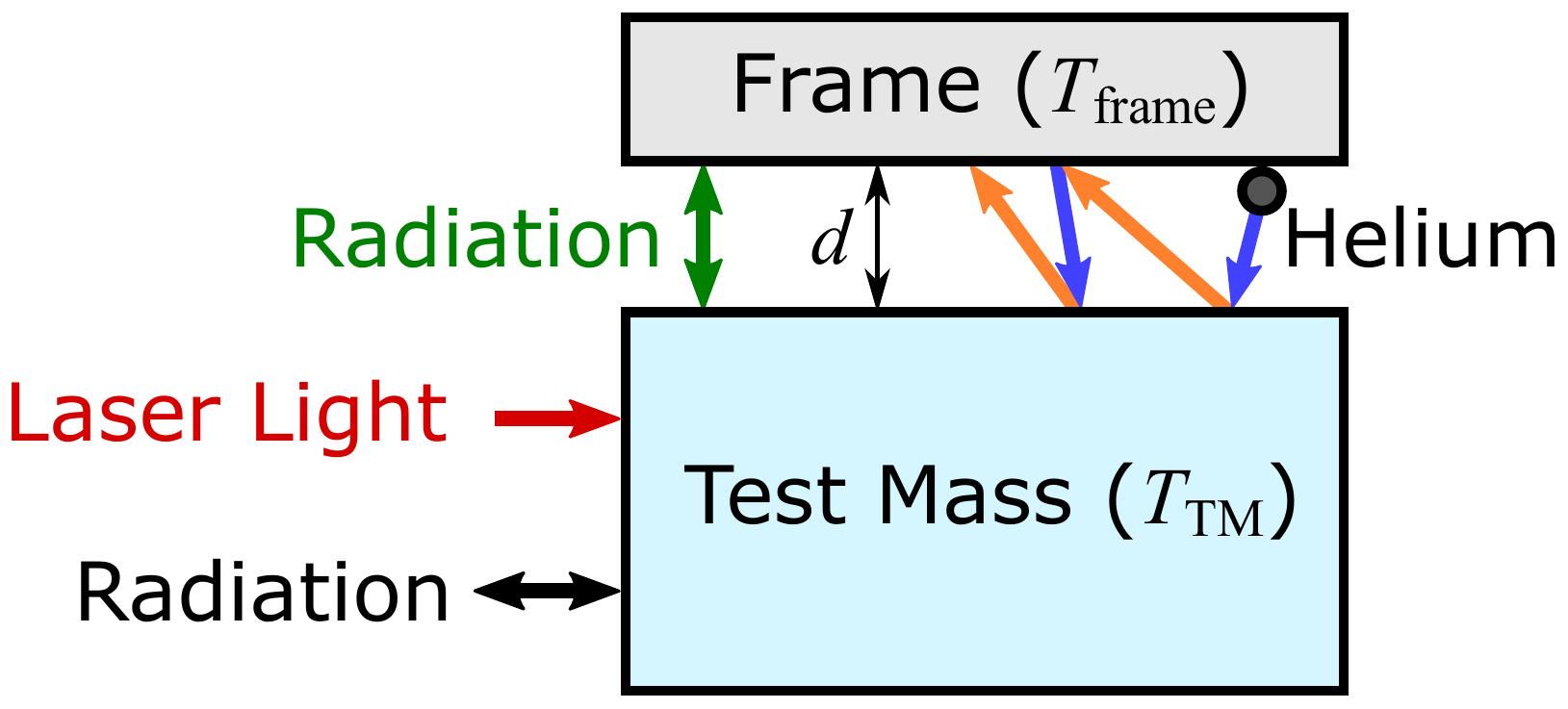}
    \caption{
        Schematic of a heat transfer model for gas cooling of a suspended test mass (TM).
        Partial absorption of laser light and thermal radiation from the environment heat the TM.
        The TM's temperature $T_{\mathrm{TM}}$ is kept constant by virtue of heat transfer to a frame, at distance $d$ and temperature $T_\mathrm{frame}$, via helium gas and thermal radiation.
        Gas cooling occurs in the free molecular flow regime (no interaction between helium atoms).
        Furthermore, we assume that helium atoms reach thermal equilibrium with mirror and frame.}
    \label{fig:suspended-mirror-setup}
\end{figure}

To establish models for cooling power and corresponding thermal displacement noise related to helium gas interacting with a suspended test mass (TM), we consider the setup shown in~\autoref{fig:suspended-mirror-setup}.
Here, a TM suspended by thin fibres (not shown) is heated by partial absorption of laser light and thermal radiation.
The imparted heat is transferred to a close-by frame, with temperature $T_\mathrm{frame}$, by virtue of helium gas and thermal radiation, thereby keeping the TM's temperature $T_{\mathrm{TM}}$ constant.
Thermal conduction through suspension fibres is excluded from the models presented.

The following underlying assumptions are made for the interaction between helium gas and surfaces of TM and frame:
The helium gas is in the free molecular flow regime, where interactions between atoms are negligible (see~\autoref{subsec:combined-model} for a discussion of potential consequences from gradually ``turning on'' mutual collisions between helium atoms);
we apply the common definition for free molecular flow based on the Knudsen number: $\mathrm{Kn}\equiv\lambda/d>10$, with mean free path (MFP) $\lambda$ and distance $d$ between TM and frame.
The MFP corresponds to the average distance traveled by atoms between successive collisions.
In~\autoref{sec:cooling-power-and-added-thermal-noise-for-et-design} it is shown, that the assumption of free molecular flow leads to technically feasible values of $d$.

The transfer of heat between helium atoms and surfaces is specified by a corresponding accommodation coefficient $\alpha_E$, which represents the fraction of incident atoms reaching thermal equilibrium with the surface~\cite{corruccini1959gaseous}.
Throughout this article, we consider $T_\mathrm{frame}=5\,\mathrm{K}$ and $T_\mathrm{TM}=18\,\mathrm{K}$, with accommodation coefficients for helium $\alpha_E\left(5\,\mathrm{K}\right)=1.0$, $\alpha_E\left(18\,\mathrm{K}\right)\equiv\alpha_{E,\mathrm{TM}}=0.6$~\cite{corruccini1959gaseous}.
As a consequence, all gas particles emitted from the frame are assumed to be in thermal equilibrium with the frame.
For the gas particles emitted from the TM, we assume that a fraction given by $\alpha_{E,\mathrm{TM}}$ is in thermal equilibrium with the TM\@.
The fraction given by $1-\alpha_{E,\mathrm{TM}}$ remains in thermal equilibrium with the frame, upon reflection from the TM\@.
Expressing heat exchange between gas particles and surfaces in terms of the thermal accommodation coefficient is a common approach, which gives good agreement with experimental results (see, e.g.~\cite{trott2011experimental}).
Similarly, the transfer of momentum between helium atoms and surfaces is specified by an accommodation coefficient $\alpha_M$, which represents the fraction of incident atoms transferring their momentum to the surface.
Here, we assume $\alpha_M\left(5\,\mathrm{K}\right)=\alpha_M\left(18\,\mathrm{K}\right)=1.0$~\cite{agrawal2008survey, cao2005temperature}.
Note that unity accommodation for momentum corresponds to the worst case in terms of frictional force acting on the TM\@.
Gas particles, which reach thermal equilibrium with or transfer their momentum to a surface, upon reflection from this surface, are said to be diffusely reflected, where the direction of re-emission from the surface is randomly distributed according to the Knudsen cosine law (see~\autoref{subsec:heat-transfer-model} for details).
The fractions of atoms $1-\alpha_E$ and $1-\alpha_M$, being reflected without exchange of energy or momentum with the surface, are said to be specularly reflected.
We assume the duration between adsorption and desorption of an atom to be negligible.

With regard to the geometry of the frame, we assume that it fully encloses the TM's cylinder barrel (in the sense, that all lines of sight departing from the barrel reach the frame) without touching it.
The underlying assumptions are consistent with a cylindrical tube concentrically surrounding a cylindrical TM at distance $d$.
This tube would have to comprize circular baffles at its ends, e.g., with an outer diameter equal to the tube's diameter and an inner diameter marginally larger than the TM's diameter to prevent touching it.
In this case, the effect on cooling power and residual gas damping noise, from not perfectly confining the gas between TM and frame, is assumed to be insignificant (see also next paragraph).
With regard to a practical implementation, one may envisage a setup similar to, e.g., the Advanced Virgo payload~\cite{naticchioni2018payloads}, with a modified reaction cage to accommodate cryogenic cooling and a baffle encapsulating the TM's barrel.

For the imperfect confinement of helium gas between TM and frame, we assume that gas particles leaking out of this region will be efficiently deflected and pumped (e.g., via cryopumping~\cite{gil2009cryogenic}), such that only an insignificant fraction of gas particles will be reflected back onto the TM\@.
As a consequence, transfer of heat or seismic/acoustic noise to the TM from other structures inside a cryostat (see e.g.~\cite{abernathy2011einstein}) is neglected.
The impact from gas damping noise, caused by helium atoms impinging on the front and backside of the TM, is considered to be insignificant, under the same assumption.
An additional noise source originates from a time-varying refractive index, arising as a consequence of a fluctuating number of gas particles traversing the optical beam path inside the interferometer~\cite{zucker1996measurement}.
For helium gas leaking out of the frame, we consider this noise to be insignificant, under the assumption that the partial pressure of the helium gas will be lowered to $\sim10^{-8}\,\mathrm{Pa}$ in the vicinity of the TM\@.
One may envisage that a corresponding cryopump could be incorporated in a helium-cooled section of the beam tube, directly adjacent to the TM cryostat.
Such a cold beam tube section is described, e.g., in~\cite{abernathy2011einstein}.

The analytical models and numerical simulations, presented in the following section, are based on 3-dimensional gaskinetic theory;
no simplification regarding the dimensionality is made.
The setup studied here is simplified with regard to a possible practical implementation, as we only consider a single pendulum, representing the suspended TM, and omit other elements of a possible payload, such as the suspended marionette (see, e.g.,~\cite{abernathy2011einstein}).
The motivation behind this simplification is, that only the noise acting on the suspended TM is affected by gas cooling.
This approach is consistent with~\cite{ET2020einstein} (Sec.~6.10.2.4), where for the suspension thermal noise, modified as a result of changing parameters of the TM's suspension fibres, with respect to the original ET design~\cite{abernathy2011einstein}, just the contribution from the suspended TM is assessed.

\section{Models for heat transfer and strain noise}
\label{sec:models-for-heat-transfer-and-strain-noise}
\subsection{Heat transfer model}
\label{subsec:heat-transfer-model}

\subsubsection{General model} \hfill\\
The cooling power acting on the TM due to the helium gas is given by
\begin{eqnarray}
    P_{\mathrm{gas}}=\alpha_{E,\mathrm{TM}}\left(\dot{Q}_{\mathrm{out}}-\dot{Q}_{\mathrm{in}}\right),
    \label{eq:cooling_power}
\end{eqnarray}
where $\dot{Q}_{\mathrm{out}}$ $\left(\dot{Q}_{\mathrm{in}}\right)$ is the heat flux emitted (adsorbed) by the TM, in the case of unity accommodation, and $\alpha_{E,\mathrm{TM}}$ represents the fraction of incident atoms contributing to the heat transfer (see~\autoref{sec:suspended-mirror-setup}).
The heat flux per surface element $\Delta A$ is calculated as the product of the flux density of emitted (adsorbed) atoms $\phi_{\mathrm{out}}$ ($\phi_{\mathrm{in}}$) and their mean kinetic energy,
\begin{eqnarray}
    \frac{\dot{Q}_{\mathrm{i}}}{\Delta A} = \phi_{i}\int\frac{1}{2}m_{\mathrm{He}} v_i^2 \rho \left(v_{i}\right) dv_{i} \varrho \left(\theta_{i}\right) d\Omega_i,
    \label{eq:d_heat}
\end{eqnarray}
with $i\in\left\{\mathrm{out},\mathrm{in}\right\}$, helium mass $m_{\mathrm{He}}$, speed of emitted atoms $v_{i}$, speed distribution $\rho \left(v_{i}\right)$, angle between the surface normal and speed vector of atom $\theta_{i}$, angular distribution $\varrho \left(\theta_{i}\right)$, and incremental solid angle $d\Omega_i=\sin\theta_{i}d\theta_{i}d\varphi_{i}$, where $\varphi_{i}$ is the azimuthal angle.
The assumptions underlying~\autoref{eq:d_heat} are described in~\autoref{sec:suspended-mirror-setup}.
The corresponding speed and angular distributions are given by~\cite{celestini2008cosine}
\begin{eqnarray}
    \rho \left(v_{i}\right)=\frac{v^3_i}{2v_{T,i}^4}\exp\left(-\frac{v^2_i}{2v_{T,i}^2}\right)
    \label{eq:speed}
\end{eqnarray}
and
\begin{eqnarray}
    \varrho \left(\theta_{i}\right)=\frac{\cos\theta_{i}}{\pi},
    \label{eq:angle}
\end{eqnarray}
respectively.
Here $v_{T,i}\equiv\sqrt{k_{\mathrm{B}} T_{i}/m_{\mathrm{He}}}$ is the characteristic thermal velocity, with $T_{\mathrm{in}} = T_{\mathrm{frame}}$ and $T_{\mathrm{out}} = T_{\mathrm{TM}}$.
The contribution from temperature-induced position fluctuations of the TM to the relative speed between TM and gas particles is insignificant and, therefore, is not taken into account.

\subsubsection{Solution for parallel plates}\hfill\\
\label{sec:parallel_plates}
We consider the simple case of two parallel plates, for which~\autoref{eq:d_heat} gives $\dot{Q}_{i}/\Delta A = 2k_{\mathrm{B}} T_{i} \phi_{i}$.
This expression is obtained by substituting the 3-dimensional distributions~\autoref{eq:speed} and~\autoref{eq:angle} into~\autoref{eq:d_heat} and integrating $v_i$ on $\left[0,\infty\right]$, $\theta_i$ on $\left[0,\pi/2\right]$, and $\varphi_i$ on $\left[0,2\pi\right]$.
These limits of integration correspond to the entire hemisphere above the surface element.
Note that the particular shape of the frame (see~\autoref{fig:suspended-mirror-setup}) is irrelevant, as long as it covers the entire hemisphere on each surface element of the TM\@.
This is because the Knudsen cosine law~\autoref{eq:angle} implies, that the number of gas particles, emitted from a surface element per solid angle into a particular direction, is independent of the orientation of the surface element with respect to the direction of emission.
The reasoning behind this is, that the solid angle subtended by a surface element on the frame, with respect to a point on the TM, is proportional to the cosine of the angle between the normal direction of this surface element and the line connecting it to the point on the TM\@.
Based on the Knudsen cosine law, the same proportionality applies to the number of molecules emitted in this particular direction.
As a result, only the total solid angle subtended by the frame matters, its shape is irrelevant, as long as its separation from the TM is compatible with the free molecular flow (see~\autoref{sec:suspended-mirror-setup}).
Taking into account equilibrium conditions, with equal incoming and outgoing flux $\phi_{\mathrm{out}}=\phi_{\mathrm{in}}\equiv\phi$, gives
\begin{eqnarray}
    \frac{\dot{Q}_{i}}{\Delta A} = 2k_{\mathrm{B}} T_{i}\phi.
    \label{eq:heat}
\end{eqnarray}

As a next step, the relation between heat transfer and helium pressure is established:
The pressure caused by incoming/outgoing atoms is given by the product of $\phi_i$ and the mean momentum along the surface normal
\begin{eqnarray}
    p_{\mathrm{i}} = \phi_{i} \int m_{\mathrm{He}} v_{i\perp} \rho \left(v_{i}\right) dv_{i} \varrho \left(\theta_{i}\right) d\Omega_{i},
    \label{eq:pressure_integral}
\end{eqnarray}
with $v_{i\perp}=v_{i} \cos \theta_{i}$.
Evaluating the integral for two parallel plates gives
\begin{eqnarray}
    p_{\mathrm{i}} = \phi \sqrt{\frac{\pi m_{\mathrm{He}} k_{\mathrm{B}} T_{i}}{2}}.
    \label{eq:pressure}
\end{eqnarray}
Gas components coming from the frame and TM have different pressures resulting from different temperatures.
This is a consequence of the assumptions detailed in~\autoref{sec:suspended-mirror-setup} (i.e., helium atoms do not interact with each other and reach thermal equilibrium with surfaces of TM and frame).
By combining~\autoref{eq:cooling_power},~\autoref{eq:heat}, and~\autoref{eq:pressure} the cooling, acting on the plate at $T_\mathrm{TM}$, can be written\footnote{There is an additional contribution from internal degrees of freedom, adding $\sim0.8~\%$ of cooling power (see, e.g.~\cite{gombosi1994gaskinetic}), which is not taken into account here.}
\begin{eqnarray}
    P_{\mathrm{gas}}=\alpha_{E,\mathrm{TM}}\sqrt{\frac{8k_{\mathrm{B}}}{\pi m_{\mathrm{He}}T_{\mathrm{frame}}}} p_{\mathrm{in}} \Delta A \left(T_{\mathrm{TM}}-T_{\mathrm{frame}}\right), && \left(T_{\mathrm{TM}}>T_{\mathrm{frame}}\right),
    \label{eq:cooling_power2}
\end{eqnarray}
where $\Delta A$ corresponds to the surface area of each plate.
Based on the reasoning provided in the paragraph before~\autoref{eq:heat},~\autoref{eq:cooling_power2} also describes the heat transfer between two concentric cylinders, where $\Delta A$ corresponds to the surface area of the inner cylinder~\cite{corruccini1959gaseous}.

\subsubsection{Maximum allowed distance between test mass and frame} \hfill\\
In the following, the maximum allowed value for the distance $d$ between frame and TM (see~\autoref{fig:suspended-mirror-setup}) is derived.
The upper bound follows from the requirement of staying in the free molecular flow regime, where $d<\lambda/10$ (see~\autoref{sec:suspended-mirror-setup}).
The total MFP corresponds to the average of the MFP of ``cold'' atoms moving from the frame toward TM,  $\lambda_{\mathrm{in}}$, and the MFP of ``hot'' (``cold'') particles moving from TM toward frame, $\lambda_{\mathrm{out_1}}$ $\left(\lambda_{\mathrm{out_2}}\right)$:
\begin{equation}
    \lambda=\frac{n_{\mathrm{in}}\lambda_{\mathrm{in}}+n_{\mathrm{out_1}}\lambda_{\mathrm{out_1}}+n_{\mathrm{out_2}}\lambda_{\mathrm{out_2}}}{n_{\mathrm{in}}+n_{\mathrm{out_1}}+n_{\mathrm{out_2}}},
    \label{eq:mfp}
\end{equation}
where $n_{\mathrm{in}}$ and $n_{\mathrm{out_1}}$ $\left(n_{\mathrm{out_2}}\right)$ are the number densities of incoming and outgoing ``warm'' (``cold'') atoms, respectively.
Here, the splitting in ``warm'' and ``cold'' particles, moving from the TM toward the frame, is a consequence of $\alpha_{E,\mathrm{TM}}<1$.
Combining~\autoref{eq:pressure} with the ideal gas law $p_{i}=n_i k_{\mathrm{B}}T_{i}$, gives:
\begin{eqnarray}
    n_\mathrm{in}&=\phi\sqrt{\frac{\pi m_\mathrm{He}}{2k_\mathrm{B} T_\mathrm{frame}}} \\
    n_\mathrm{out,1}&=\alpha_{E,\mathrm{TM}}\phi\sqrt{\frac{\pi m_\mathrm{He}}{2k_\mathrm{B} T_\mathrm{TM}}} \\
    n_\mathrm{out,2}&=\left(1-\alpha_{E,\mathrm{TM}}\right)\phi\sqrt{\frac{\pi m_\mathrm{He}}{2k_\mathrm{B} T_\mathrm{frame}}}.
    \label{eq:ns}
\end{eqnarray}
Substituting these expressions into~\autoref{eq:mfp} yields
\begin{eqnarray}
    \lambda=\frac{\lambda_{\mathrm{in}}+\alpha_{E,\mathrm{TM}}\sqrt{T_{\mathrm{frame}}/T_{\mathrm{TM}}}\lambda_{\mathrm{out_1}}+\left(1-\alpha_{E,\mathrm{TM}}\right)\lambda_{\mathrm{out_2}}}{2+\alpha_{E,\mathrm{TM}}\left(\sqrt{T_{\mathrm{frame}}/T_{\mathrm{TM}}}-1\right)}.
    \label{eq:mfp_total}
\end{eqnarray}
We follow the common approach for deriving the MFP, as, for example, presented in~\cite{jeans_1982}: The average distance travelled by a gas atom of ``species'' $i\in\left\{\mathrm{in},\mathrm{out_1},\mathrm{out_2}\right\}$ between two successive collisions (i.e., the MFP) is given by $\lambda_i=\langle v_i\rangle \tau_i$, where $\langle v_i\rangle$ is the mean speed and $\tau_i$ is the average time between successive collisions.
Atoms can collide either with atoms of their own species, characterized by collision time $\tau_{i,i}$, or atoms of the other species, characterized by collision times $\tau_{i,j}$ $\left(i\neq j\right)$.
The total collision time for a particular species $i$ is calculated by summing the contributing collision rates: ${\tau_i}^{-1}=\sum_j\tau_{i,j}^{-1}$.
The rate of each collision processes is calculated by multiplying the volume of interaction per unit time by the number density of target gas particles: $\tau_{i,j}^{-1}=\pi \delta^2 n_j \langle v_{i,j} \rangle$, where $\delta$ is the kinetic diameter of a Helium atom and $\langle v_{i,j} \rangle$ is the mean relative speed between an atom of species $i$ and an atom of species $j$.
Combining the previous considerations yields
\begin{eqnarray}
    \fl\lambda_{\mathrm{in}}&=\frac{k_{\mathrm{B}}T_{\mathrm{frame}}\langle v_{\mathrm{in}}\rangle}{\pi \delta^2 p_{\mathrm{in}} \left[\langle v_{\mathrm{in,in}} \rangle+\alpha_{E,\mathrm{TM}}\sqrt{T_{\mathrm{frame}}/T_{\mathrm{TM}}}\langle v_{\mathrm{in,out_1}} \rangle + \left(1-\alpha_{E,\mathrm{TM}}\right)\langle v_{\mathrm{in,out_2}} \rangle \right]}, \label{eq:mfp_contributors_one} \\
    \fl\lambda_{\mathrm{out_1}}&=\frac{k_{\mathrm{B}}T_{\mathrm{TM}}\langle v_{\mathrm{out_1}}\rangle}{\pi \delta^2 p_{\mathrm{in}} \left[\alpha_{E,\mathrm{TM}}\sqrt{T_{\mathrm{frame}}/T_{\mathrm{TM}}}\langle v_{\mathrm{out_1,out_1}} \rangle+\langle v_{\mathrm{out_1,in}} \rangle + \left(1-\alpha_{E,\mathrm{TM}}\right) \langle v_{\mathrm{out_1,out_2}} \rangle \right]}, \\
    \fl\lambda_{\mathrm{out_2}}&=\frac{k_{\mathrm{B}}T_{\mathrm{frame}}\langle v_{\mathrm{out_2}}\rangle}{\pi \delta^2 p_{\mathrm{in}} \left[\left(1-\alpha_{E,\mathrm{TM}}\right) \langle v_{\mathrm{out_2,out_2}} \rangle+\langle v_{\mathrm{out_2,in}} \rangle + \alpha_{E,\mathrm{TM}}\sqrt{T_{\mathrm{frame}}/T_{\mathrm{TM}}}\langle v_{\mathrm{out_2,out_1}} \rangle \right]}.
    \label{eq:mfp_contributors}
\end{eqnarray}
Here, the mean speed is given by
\begin{eqnarray}
    \langle v_{i}\rangle=\sqrt{\frac{9\pi k_{\mathrm{B}} T_{i}}{8m_{\mathrm{He}}}},
    \label{eq:avg_speed}
\end{eqnarray}
with $T_\mathrm{in}=T_\mathrm{out_2}=T_\mathrm{frame}$ and $T_\mathrm{out_1}=T_\mathrm{TM}$, and the mean relative speed, defined as $\langle v_{i,j} \rangle=\sqrt{\langle \left(\vec{v}_i-\vec{v}_j\right)^2\rangle}=\sqrt{\langle v_i\rangle^2 + \langle v_i\rangle^2-2\langle \vec{v}_i\cdot\vec{v}_j\rangle^2}$, is given by
\begin{eqnarray}
    \langle v_{\mathrm{in,in}} \rangle=\langle v_{\mathrm{out_2,out_2}} \rangle &= \sqrt{\frac{k_{\mathrm{B}}T_\mathrm{frame}}{m_{\mathrm{He}}}\left(8-\pi\right)} \\
    \langle v_{\mathrm{out_1,out_1}} \rangle &= \sqrt{\frac{k_{\mathrm{B}}T_\mathrm{TM}}{m_{\mathrm{He}}}\left(8-\pi\right)} \\
    \langle v_{\mathrm{out_1,out_2}} \rangle=\langle v_{\mathrm{out_2,out_1}} \rangle &= \sqrt{\frac{k_{\mathrm{B}}}{m_{\mathrm{He}}}\left[4\left(T_{\mathrm{frame}}+T_{\mathrm{TM}}\right)-\pi\sqrt{T_{\mathrm{frame}}T_{\mathrm{TM}}}\right]} \\
    \langle v_{\mathrm{in,out_1}} \rangle=\langle v_{\mathrm{out_1,in}} \rangle &= \sqrt{\frac{k_{\mathrm{B}}}{m_{\mathrm{He}}}\left[4\left(T_{\mathrm{frame}}+T_{\mathrm{TM}}\right)+\pi\sqrt{T_{\mathrm{frame}}T_{\mathrm{TM}}}\right]} \\
    \langle v_{\mathrm{in,out_2}} \rangle=\langle v_{\mathrm{out_2,in}} \rangle &= \sqrt{\frac{k_{\mathrm{B}}T_\mathrm{frame}}{m_{\mathrm{He}}}\left(8+\pi\right)}.
    \label{eq:avg_rel_speed}
\end{eqnarray}
The mean values are calculated based on the distribution functions given by~\autoref{eq:speed} and~\autoref{eq:angle}.

Substituting~\autoref{eq:mfp_contributors_one} to~\autoref{eq:avg_rel_speed} into~\autoref{eq:mfp_total} enables calculating $\lambda$ based on the parameters given in~\autoref{sec:suspended-mirror-setup}.
The upper bound for the maximum distance between frame and TM, which is compatible with free molecular flow, is given by $d=\lambda/10$.

\subsubsection{Comparison between analytical and numerical models} 
\begin{figure}
    \centering
    \includegraphics[width=0.7\textwidth]{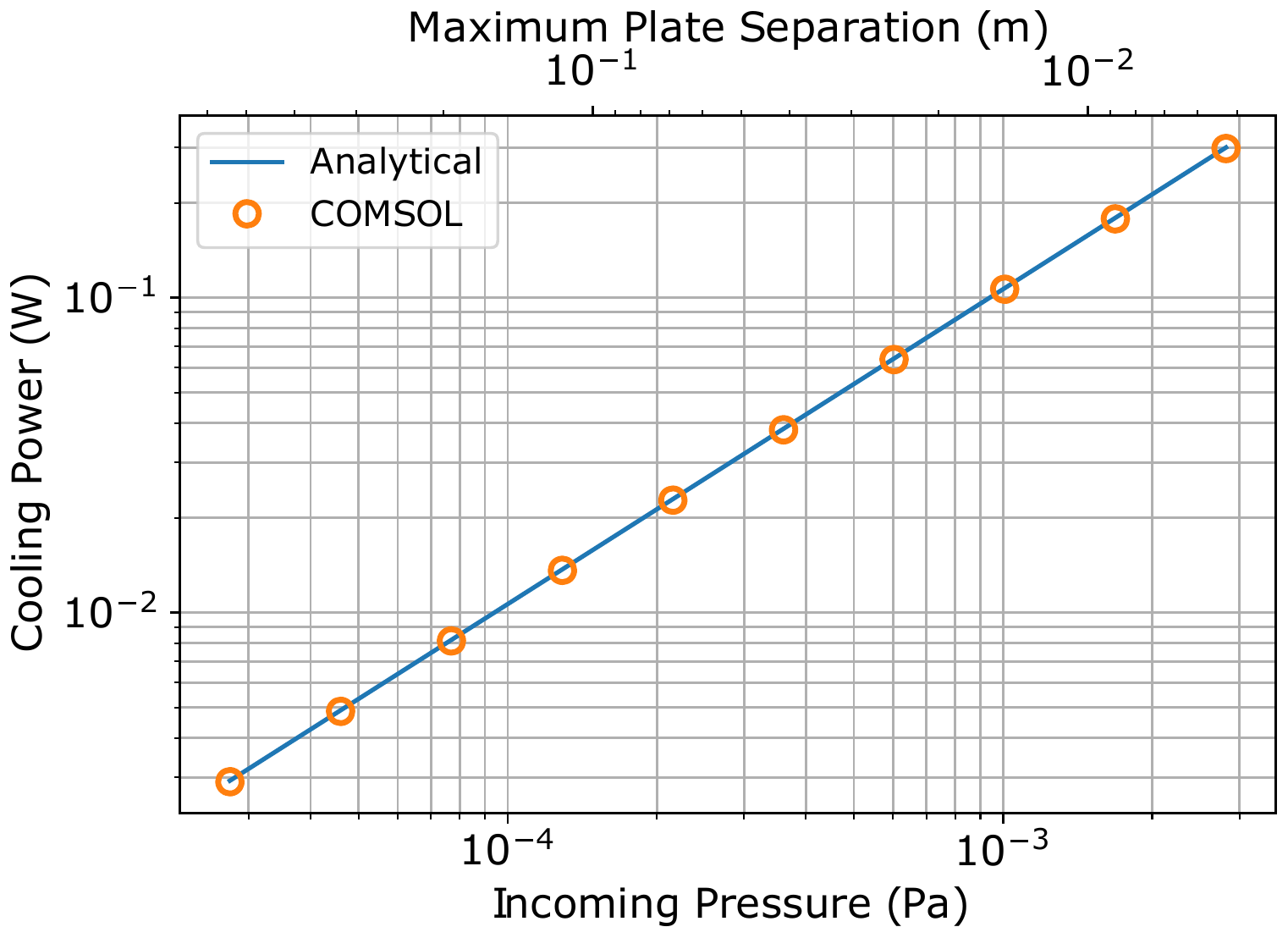}
    \caption{
        Heat transfer versus helium gas pressure for two parallel plates of area $\SI{50}{cm} \times \SI{50}{cm}$, unity accommodation, and temperatures \SI{5}{K} and \SI{18}{K}, respectively.
        The blue line shows the analytical model~\autoref{eq:cooling_power2}.
        Values from the numerical simulation (orange circles) are lower by 0.02~\%.
        The upper x-axis indicates the separation between the plates, where shown values correspond to the upper bound compatible with free molecular flow (see main text for details).}
    \label{fig:heat-transfer-models}
\end{figure}

\hfill\\
To validate the model presented in the previous sections, we compare it with a numerical simulation implemented in the Molecular Flow Module of COMSOL Multiphysics.
We put two plates inside a closed box, which is consistent with the analytical calculation in the sense that the entire hemisphere above the warm plate is covered (see~\autoref{sec:parallel_plates}).
Figure~\ref{fig:heat-transfer-models} shows the predicted heat transfer as a function of pressure $p_1$ for a pair of parallel square plates, with edge length \SI{50}{cm} and unity accommodation at both plates.
The blue line corresponds to the analytical expression~\autoref{eq:cooling_power2} and the orange circles show the values predicted by the numerical simulation.
The analytical values exceed the numerical values by 0.02~\%.
Increasing the pressure lowers the MFP, which requires a reduced separation between the plates, to be compatible with the free molecular flow regime ($\mathrm{Kn}>10$).
The upper x-axis shows the maximum separation between the plates still compatible with the free molecular flow, based on~\autoref{eq:mfp_total}.

\subsection{Strain noise model}
\label{subsec:strain-noise-model}
Here, we derive the displacement noise arising from helium atoms impinging on the TM~\cite{cavalleri2010gas}.
As dominant noise source we consider a single degree of freedom of the TM, corresponding to motion along the direction of the incident laser light (see~\autoref{fig:suspended-mirror-setup}) with velocity $V_{\parallel}$.

In addition to impinging gas atoms, diffusive gas flow is a potential source of noise acting on a TM in a constrained volume.
The corresponding noise becomes significant if the channel limiting the flow (i.e.\ a gap between the TM and a nearby surface) is comparable to the dimension of the TM~\cite{schlamminger2010comparison,cavalleri2009increased,weiss2009gas}.
Here, we assume that gas can flow around the frame in a channel larger than TM and frame.
In this case, the gap between TM and frame (see~\autoref{fig:suspended-mirror-setup}) is not limiting the flow resulting from pressure differences caused by the thermal motion of the TM along the optical axis.
Therefore, we consider the noise contribution from diffusive flow negligible.

The time-averaged force per surface element $\Delta A$ of the TM's side faces is calculated as the mean value of the gas particles' flux density of momentum parallel to the TM motion
\begin{eqnarray}
  \frac{F_{\parallel}}{\Delta A} = \phi_{\mathrm{in}} \int m_{\mathrm{He}} v_{\mathrm{in},\parallel} \rho_{V_{\parallel}} \left(v_{\mathrm{in}}\right) dv_{\mathrm{in}} \varrho \left(\theta_{\mathrm{in}}\right) d\Omega_{\mathrm{in}}, 
  \label{eq:force_parallel_integral}	
\end{eqnarray} 
with speed distribution 
\begin{eqnarray}
	\rho_{V_{\parallel}} \left(v_{\mathrm{in}}\right)= \frac{v^3_{\mathrm{in}}}{2 v_T^4} \exp \left[-\frac{v^2_{\mathrm{in},\perp}+\left(v_{\mathrm{in},\parallel}+V_{\parallel}\right)^2}{2 v_T^2}\right].
  \label{eq:speed_parallel}	
\end{eqnarray}
Here, $v_{\mathrm{in},\perp}=v_{\mathrm{in}}\cos\theta_{\mathrm{in}}\vec{e}_z$ and $v_{\mathrm{in},\parallel}=v_{\mathrm{in}}\sin \theta_{\mathrm{in}}\left(\cos\varphi_{\mathrm{in}}\vec{e}_x+\sin\varphi_{\mathrm{in}}\vec{e}_y\right)$ are the gas particles' speed components orthogonal and parallel to $V_{\parallel}$, respectively, where $\varphi_{\mathrm{in}}$ is the azimuthal angle. 
Solving the integral gives
\begin{eqnarray}
  F_{\parallel} =-V_{\parallel} p_{\mathrm{in}} \Delta A \sqrt{\frac{2m_{\mathrm{He}}}{\pi k_{\mathrm{B}}T_{\mathrm{frame}}}} \equiv - V_{\parallel}\beta,
  \label{eq:force_parallel}	
\end{eqnarray}
where the last step defines the damping coefficient $\beta$.
Note that atoms re-emitted from the TM do not cause a net force. 
This is because emission occurs isotropically. 
The integral is solved analogously to the one for the heat transfer model.
The details are discussed in paragraph 2 of~\autoref{subsec:heat-transfer-model}.
\autoref{eq:force_parallel} is applicable to TMs in the shape of cylinders or cubes (or any right prism) oscillating along the direction defined by a surface normal of a side face~\cite{cavalleri2010gas}.
$\Delta A$ corresponds to the surface area of the cylinder's barrel or four of the cube's side faces, respectively.

Assuming the TM to represent a damped harmonic oscillator, with frictional damping force $F_{\parallel}$, and applying the fluctuation-dissipation theorem yields the displacement noise spectrum~\cite{saulson1990thermal}
\begin{eqnarray}
	x^2\left(\omega\right)=\frac{4k_{\mathrm{B}}T_\mathrm{frame}\beta}{m^2_{\mathrm{TM}}\left(\omega_0^2-\omega^2\right)^2+\beta^2\omega^2},
	\label{eq:disp_noise}	
\end{eqnarray}
where $m_{\mathrm{TM}}$ is the mass of the TM, $\omega_0/2\pi$ is the oscillator's resonance frequency, and $\omega/2\pi$ is the frequency. 

To validate our analytical model, we set up a Monte-Carlo simulation~\cite{evans2011gas}.
Here, the suspended TM is modeled as a pendulum (using a small-angle approximation), with length $l$ and  displacement $z(t)$.
Further assumptions are described in~\autoref{sec:suspended-mirror-setup}.
For each impinging atom, we assign a random timestamp and calculate the corresponding momentum transfer, based on randomly selecting four parameters: in- and outgoing speed as well as in- and outgoing angle.
The underlying probability distributions for speed and angle are given by~\autoref{eq:speed} and~\autoref{eq:angle}, respectively.
The change in the TM's velocity associated with the momentum transferred by an adsorbed gas atom is given by
\begin{equation}
    \delta V_\parallel =
    \frac{m_\text{He}}{m_\text{TM}}
    \left[
    v(T_\text{frame}) \cdot
    \sin \theta_\text{in} +
    v(T_\text{TM}) \cdot
    \sin \theta_\text{out}
    \right].
    \label{eq:velocity_distribution}
\end{equation}
The TM's trajectory is obtained based on energy conservation, giving:
\begin{equation}
    z(t) \approx l\left( \sqrt{2\xi} + \frac{1}{24}\sqrt{2\xi}^{3} \right) \sin\left(\omega_0 t + \frac{z_0}{l}\right),
\end{equation}
with acceleration due to gravity $g$,
velocity added to the TM at the last hit $\delta V_\parallel$,
TM displacement at the last collision $z_0$, and
\begin{equation}
   \xi = \frac{1}{2l} \left[\frac{z_0^2}{l}+\frac{1}{g}\left(V_\parallel+\delta V_\parallel\right)^2\right].
\end{equation}
The power spectral density of corresponding displacement noise is obtained by Fourier transforming the time series of TM displacements.

\begin{figure}
  \centering
    \includegraphics{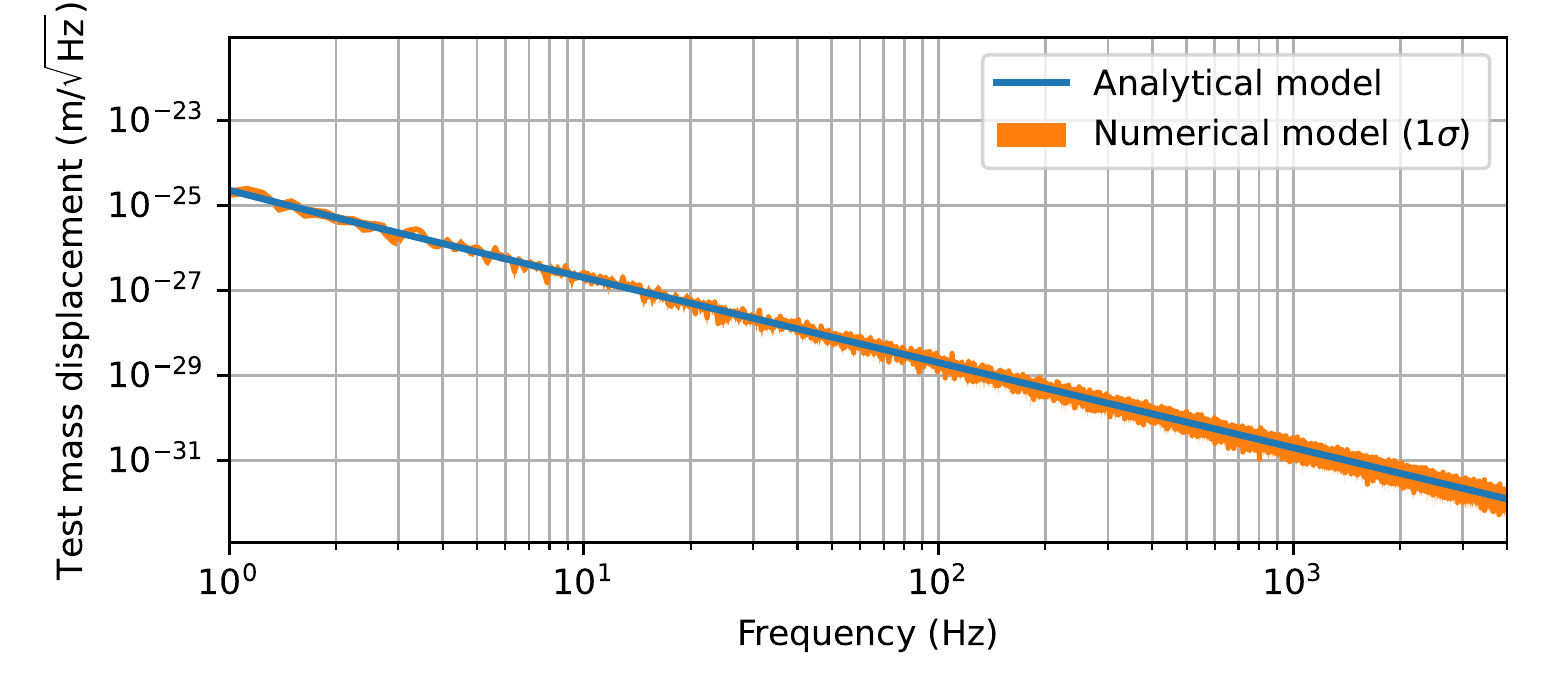}
		\caption{Residual gas damping noise from helium gas impinging on four side faces of a cubical toy model TM with mass \SI{300}{kg}, suspension length \SI{2}{m}, and rate of impinging helium atoms $1.5\times 10^{6}\,\mathrm{s^{-1}}$. (The displacement noise values shown here are of no relevance for realistic cooling scenarios, due to a significantly lower collision rate. See main text for details).
        The numerical simulation (orange) is obtained by averaging seven individual noise spectra; the width corresponds to the standard deviation.
        The analytical model~\autoref{eq:disp_noise} (consistent with~\cite{cavalleri2010gas}) is shown in blue.
				The average deviation between both models is 1~\%.
        \label{fig:psd_10M_collisions}
        }
\end{figure}

Figure~\ref{fig:psd_10M_collisions} shows the simulated displacement spectral density (orange) together with the analytical model (blue, given by~\autoref{eq:disp_noise}) for a cubical toy model TM with a mass of \SI{300}{kg}.
The average deviation between numerical simulation and analytical model is 1~\%.
The example illustrated here just serves as a comparison of the analytical and numerical model;
the resulting displacement noise values are of no relevance whatsoever for gravitational wave detectors.
With reasonable computational resources we are not able to simulate collision rates (in excess of $10^{20}\,\mathrm{s^{-1}}$) relevant for realistic cooling scenarios (see~\autoref{sec:cooling-power-and-added-thermal-noise-for-et-design}).
Therefore, in the present case, we simulate TM movement with $1.5\times 10^{6}\,\mathrm{s^{-1}}$.
Mirror movement is simulated for a time of \SI{8}{s}.
\subsection{Relation between heat transfer and strain noise}
\label{subsec:combined-model}
Combining~\autoref{eq:cooling_power2},~\autoref{eq:force_parallel}, and~\autoref{eq:disp_noise} yields the following expression (for $\omega\gg\omega_0$), directly relating the cooling power, provided by the helium atoms, to corresponding residual gas damping noise:
\begin{eqnarray}
    P_{\mathrm{gas}}=\alpha_{E,\mathrm{TM}}\frac{\omega^4 x^2\left(\omega\right)}{2N_{\mathrm{TM}}}\frac{m_{\mathrm{TM}}^2}{m_{\mathrm{He}}}\frac{\left(T_{\mathrm{TM}}-T_{\mathrm{frame}}\right)}{T_{\mathrm{frame}}},
    \label{eq:cooling_T_scaling}
\end{eqnarray}
where $N_{\mathrm{TM}}$ is the number of TMs in the detector (in all current and next generation observatories, $N_{\mathrm{TM}}=4$).
Interestingly, this expression is independent of the helium pressure $p_\mathrm{in}$ and the surface area $\Delta A$ of the TM, which is exposed to helium gas atoms.
This is a consequence of the fact that for increasing/decreasing either $p_\mathrm{in}$ or $\Delta A$, the effects from increasing/decreasing both heat flux and residual gas damping noise cancel.
For a fixed amount of added mirror displacement noise, the cooling power increases with the square of the mirror mass.
This corresponds to a quartic dependency on mirror diameter.
Increasing the temperature of the TMs with respect to the frame leads to larger cooling power for a given amount of added noise.   
In that regard, increasing the accommodation coefficient is beneficial too.
Furthermore, the cooling power for a given amount of noise increases for lighter gas atoms and lower frame temperature.
Solving~\autoref{eq:cooling_T_scaling} for the displacement noise results in an inverse proportional dependency between noise and mirror mass.
 
In the following, we consider the situation of two species of gas particles  impinging on the TM\@.
The first (second) species is characterized by flux density $\phi_1$ ($\phi_2$) and temperature $T_1=T_\mathrm{frame}$ ($T_\mathrm{frame}<T_2<T_\mathrm{TM}$).
For the cooling power acting on the TM follows, according to~\autoref{eq:cooling_power} and~\ref{eq:heat},
\begin{equation}
P_{\mathrm{gas}}=2k_\mathrm{B}\alpha_{E,\mathrm{TM}}\Delta A\left[\phi_1\left(T_\mathrm{TM}-T_\mathrm{frame}\right)+\phi_2\left(T_\mathrm{TM}-T_\mathrm{2}\right)\right].
\end{equation}
The corresponding frictional force, according to~\autoref{eq:force_parallel} and~\autoref{eq:pressure}, is given by
\begin{equation}
F_{\parallel} =-V_{\parallel} m_\mathrm{He} \Delta A \left(\phi_1+\phi_2\right).
\end{equation}
Based on the previous two equations, the frictional force per cooling power, $F_{\parallel}/P_\mathrm{gas}$, increases for an increasing fraction of ``warm'' particles, $\phi_2/\phi_1$.
As a result, the damping coefficient $\beta$ is greater, compared to the case with $\phi_2=0$.
Furthermore, calculating the TM's displacement noise spectrum in this situation requires replacing $T_\mathrm{frame}$ in~\autoref{eq:disp_noise} with an average gas temperature $T>T_\mathrm{frame}$.
Assuming this simplified situation to be representative for the aspect of increasing the average temperature of gas particles impinging on the TM, e.g., as a consequence of gradually increasing collisions between gas particles, indicates the detrimental effects of leaving the molecular flow regime towards regimes characterized by $\mathrm{Kn}<10$. 

\section{Cooling power and added thermal strain noise with regard to the ET design}
\label{sec:cooling-power-and-added-thermal-noise-for-et-design}
To assess the potential of gas cooling for future gravitational wave observatories, we examine a setup similar to the design of the cryogenic interferometer for the low-frequency ET~\cite{abernathy2011einstein,ET2020einstein}.
As for the ET design, we assume the interferometer to comprise four cylindrical TMs made out of silicon; 
each TM has a diameter $D_{\mathrm{TM}}=45\,\mathrm{cm}$, thickness $t_{\mathrm{TM}}=57\,\mathrm{cm}$, and mass $m_{\mathrm{TM}}=211\,\mathrm{kg}$.
The expected heat load from absorbed laser light and thermal radiation is $\sim100\,\mathrm{mW}$, which, according to the baseline design, is fully extracted by conduction through the TM's silicon suspension fibres.
In the case of gas cooling as complementing cooling strategy, the imparted heat is transferred to a frame surrounding the barrel of a TM (see \autoref{sec:suspended-mirror-setup}).
Here, we assume TM and frame temperatures of $T_{\mathrm{TM}}=18\,\mathrm{K}$ and $T_{\mathrm{frame}}=5\,\mathrm{K}$, respectively.
At this value for $T_{\mathrm{TM}}$, the coefficient of thermal expansion for silicon vanishes, thereby eliminating noise from thermoelastic damping.
The cooling power provided by the helium gas is calculated according to \autoref{eq:cooling_power2}, with $\Delta A=A_\mathrm{barrel}=\pi D_{\mathrm{TM}} t_{\mathrm{TM}}$.
The contribution from radiation is calculated based on the Stefan-Boltzmann law $\sigma\left(\varepsilon_\mathrm{barrel} A_\mathrm{barrel}+\varepsilon_\mathrm{face} \pi D_\mathrm{TM}^2/2\right)\left(T_{\mathrm{TM}}^4-T_{\mathrm{frame}}^4\right)$,
with Stefan-Boltzmann constant $\sigma$, emissivity of the TM's barrel $\varepsilon_\mathrm{barrel}=0.9$, and emissivity of the TM's front and backside $\varepsilon_\mathrm{TM}=0.6$~\cite{adhikari2020cryogenic}.
Here, the contribution from the TM's barrel is linear in $D_{\mathrm{TM}}$ and the contribution from the TM's side is quadratic in $D_{\mathrm{TM}}$.
For the given parameters, a radiative cooling power of \SI{5}{mW} is predicted.
\begin{figure}
    \centering
    \includegraphics[width=0.7\textwidth]{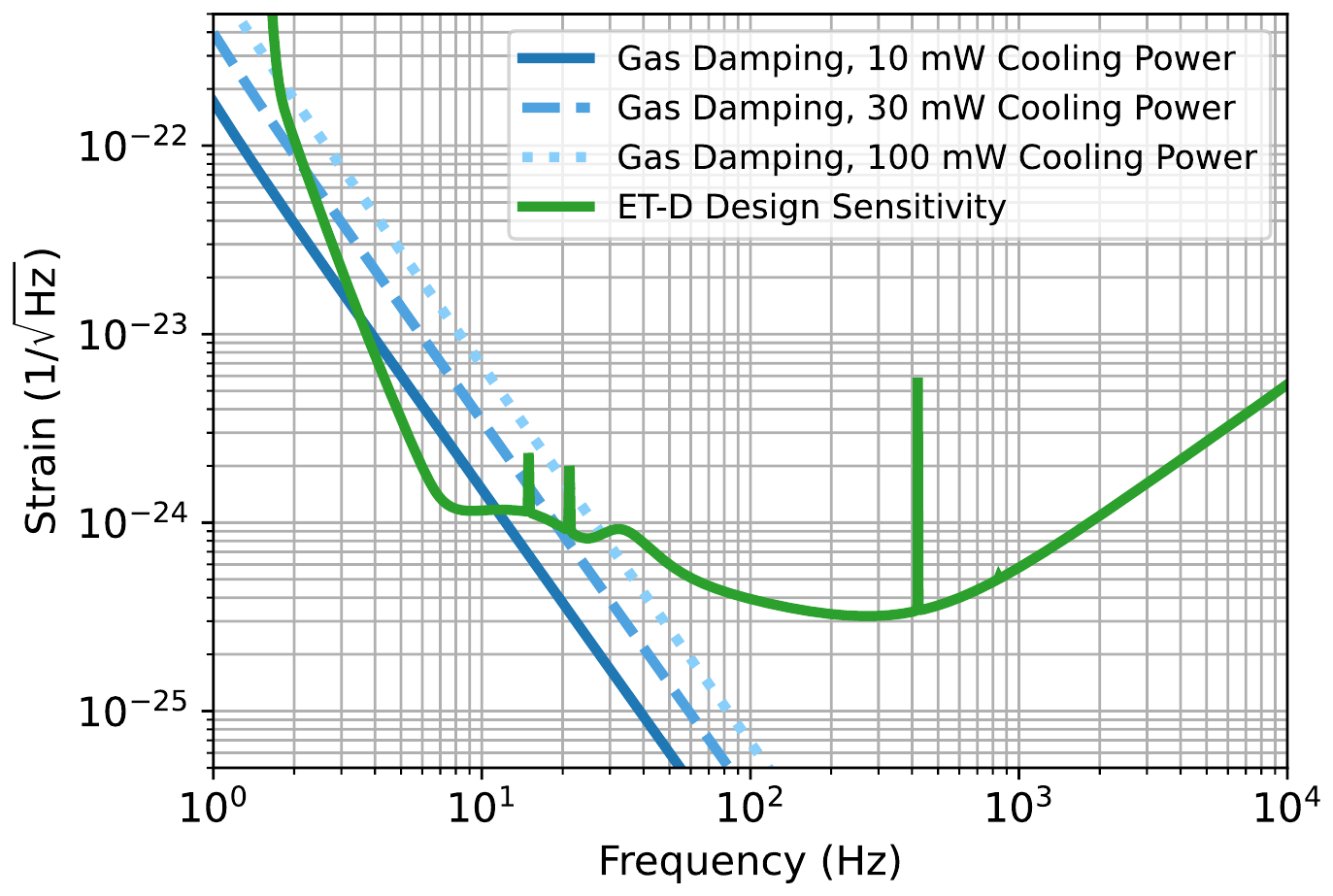}
    \caption{Residual gas damping noise from gas cooling for the Einstein Telescope design.
    The solid, dashed, and dotted blue curves show simulated strain noise spectra caused by helium atoms impinging on the four test masses, with corresponding total cooling powers per test mass of \SI{10}{mW}, \SI{30}{mW}, and \SI{100}{mW}, respectively. 
    The green curve shows the design sensitivity of the Einstein Telescope.}
    \label{fig:thermal-noise}
\end{figure}

Figure~\ref{fig:thermal-noise} shows the noise associated with gas cooling together with the sensitivity of the ET design (ET-D)~\cite{abernathy2011einstein,ET2020einstein}.
All values are given in terms of strain, which is defined as $x\left(\omega\right)/L$, where $L
$ is the distance between the two TMs in an arm of the interferometer.
Here, we assume the value of the ET design: $L=10\,\mathrm{km}$~\cite{abernathy2011einstein,ET2020einstein}.
The expected heat load on each TM is \SI{100}{mW}, according to~\cite{ET2020einstein}, Sec.~6.10.2.4.
An optimistic estimate considers just the absorption in the optical coating, contributing \SI{18}{mW} (i.e., \SI{1}{ppm} absorption of \SI{18}{kW} circulating power, according to~\cite{ET2020einstein}, Sec.~6.11.2), and an additional contribution of $\sim$~\SI{10}{mW} from thermal radiation (see~\cite{abernathy2011einstein}, Sec.~3.9.2), resulting in a corresponding heat load of $\sim$~\SI{30}{mW}.
This estimate assumes optical absorption in the substrate to be negligible.
For \SI{10}{mW} of total cooling power (e.g., a 10~\% contribution to extract the expected heat load), with equal contributions from helium gas and radiation, the noise from cooling (solid blue line) exceeds the ET-D sensitivity (green) by at most a factor of 2.3 in a narrow frequency band from 3 to 11 Hz.
The corresponding helium pressure is $2\times10^{-5}~\mathrm{Pa}$, which imposes the bound $d<66\,\mathrm{cm}$ on the distance between TM and frame, for compatibility with free molecular flow.
Extracting a heat load of \SI{30}{mW} (\SI{100}{mW}) requires \SI{25}{mW} (\SI{95}{mW}) of cooling power provided by the helium gas. 
The corresponding helium pressure is $12\times10^{-5}~\mathrm{Pa}$ $\left(46\times10^{-5}~\mathrm{Pa}\right)$, which imposes the bound $d<12\,\mathrm{cm}$ $\left(d<3\,\mathrm{cm}\right)$ for compatibility with free molecular flow.
The resulting noise, shown by the dashed (dotted) blue line exceeds the ET-D sensitivity by at most a factor of 5.3 (10.3) in the frequency band from \SI{2}{Hz} (\SI{1}{Hz}) to \SI{19}{Hz} (\SI{28}{Hz}).
These results indicate that, for the given detector configuration, increasing the cooling power comes at the cost of simultaneously increasing the residual gas damping noise.
Getting \SI{10}{mW}, \SI{30}{mW}, or \SI{100}{mW} of cooling power with a maximum added noise comparable to the ET-D sensitivity (represented by the solid green curve in \autoref{fig:thermal-noise}) requires $m_{\mathrm{TM}}\sim 500~\mathrm{kg}$, $m_{\mathrm{TM}}\sim 1100~\mathrm{kg}$, or $m_{\mathrm{TM}}\sim 2200~\mathrm{kg}$, respectively.

\begin{figure}[b]
    \centering
    \includegraphics[width=0.75\textwidth]{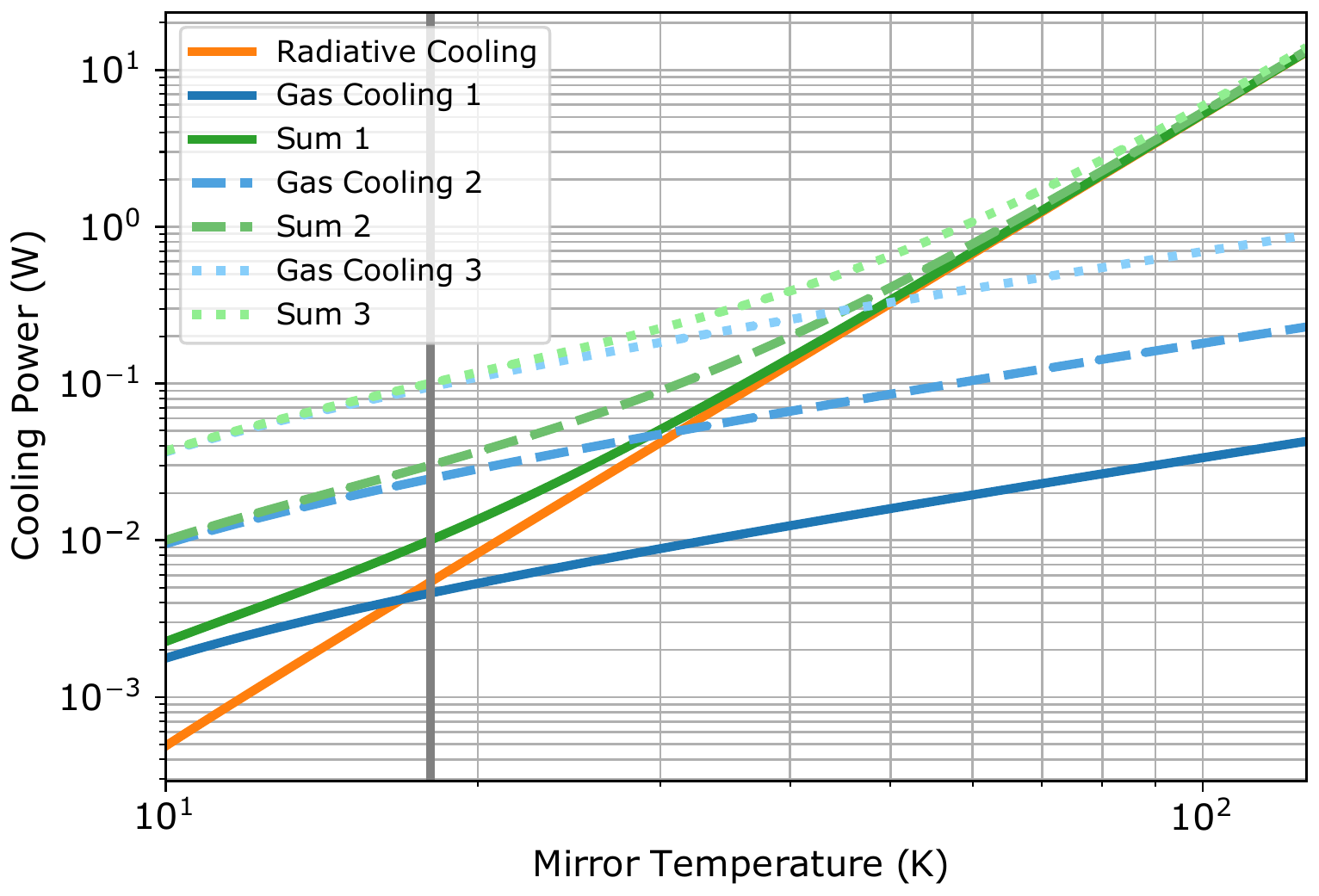}
    \caption{
    Cooling power versus test mass temperature for the Einstein Telescope design.
    The solid dashed, and dotted blue curves show the cooling power provided by helium gas according to \autoref{eq:cooling_T_scaling}, where the corresponding strain noise is shown in \autoref{fig:thermal-noise} by curves of the same formatting.
    The orange line shows the contribution from radiative heat transfer (details provided in the main text). The solid dashed, and dotted green curves show the sum of contributions from cooling by gas and radiation.
    The vertical gray line indicates a temperature of the test mass mirrors of \SI{18}{K}.}
    \label{fig:temperature_scaling}
\end{figure}

Figure~\ref{fig:temperature_scaling} shows the total cooling power and its contributors versus TM temperature.
The solid dashed, and dotted blue curves represent the contribution from helium gas, where the corresponding noise is shown by the curves of similar formatting in \autoref{fig:thermal-noise}.
For helium pressures corresponding to \SI{10}{mW}, \SI{30}{mW}, or \SI{100}{mW} of cooling power at $T_\mathrm{TM}=18\,\mathrm{K}$, radiative cooling dominates over gas cooling for $T_{\mathrm{TM}}>17\,\mathrm{K}$, $T_{\mathrm{TM}}>31\,\mathrm{K}$, or $T_{\mathrm{TM}}>50\,\mathrm{K}$, respectively.

\section{Conclusion}
\label{sec:conclusion}
Based on a conceptual setup of a suspended test mass mirror in future GW observatories, we have established a relation between gas-induced cooling power and corresponding added observatory strain noise.
In this process, we have developed analytical models for cooling power and noise, which we compared to numerical simulations, finding excellent agreement within 1~\% for both heat transfer and noise model;
our noise model is also consistent with the one presented in~\cite{cavalleri2010gas}.
For the considered setup, where heat is transferred between the mirror's cylinder barrel and a close-by frame, we have shown that the gas-induced cooling power, for a fixed amount of added mirror displacement noise, increases with the square of the mirror mass. 
This corresponds to a quartic dependency on mirror diameter for gas cooling. 
For comparison, the mirror's radiative cooling power is a sum of contributions from surfaces of cylinder barrel and sides, with linear and quadratic dependency on the mirror's diameter, respectively. 
Note that increasing mirror diameter and mass also suppresses other noise contributors and is overall beneficial for the sensitivity of gravitational wave detectors~\cite{abernathy2011einstein,ET2020einstein}.
We also have applied our theoretical framework with regard to the Einstein Telescope design, assuming a mirror temperature of \SI{18}{K} and mirror masses of \SI{211}{kg}: 
A gas cooling power of \SI{10}{mW} introduces additional noise exceeding the Einstein Telescope design sensitivity by a factor $\sim3$ in a frequency band from 3 to \SI{11}{Hz}.
A gas cooling power of \SI{100}{mW} results in additional noise exceeding the Einstein Telescope design sensitivity by a factor $\sim11$ in a frequency band from 1 to \SI{28}{Hz}.
We have pointed out that hypothetically increasing the mirror mass by a factor of 3 and 11, for a cooling power of \SI{10}{mW} and \SI{100}{mW}, respectively, reduces the gas-induced strain noise to a level, which is at most comparable to the Einstein Telescope design sensitivity. 
This illustrates the inverse proportionality between strain noise and mirror mass for a fixed gas cooling rate.
The considered cooling powers comprize an estimated \SI{5}{mW} from radiative cooling, which corresponds to 5~\% of the expected heat load.
With regard to the current baseline cooling concept for the ET project, we have pointed out the potential benefit of additional cooling power, provided by gas cooling, and resulting limitations in form of added observatory strain noise.
Gas cooling should be considered as an additional resource in combination with the currently-used conductive cooling when designing third generation detectors.
The uncertainty in the accommodation coefficient makes an experimental test of the proposed cooling approach desirable.

\section*{Acknowledgements}
We thank Sandy Croatto, Michael Hartman, and Mikhail Korobko for helpful dis\-cussions.
This work was supported and partly financed (AF) by the DFG under Germany's Excellence Strategy EXC 2121 "Quantum Universe" -- 390833306.
\section*{References}
\bibliographystyle{unsrt}
\bibliography{ref/references_gas_cooling}

\begin{thebibliography}{10}

\bibitem{abbott2016observation}
Benjamin~P Abbott, Richard Abbott, TD~Abbott, MR~Abernathy, Fausto Acernese,
  Kendall Ackley, Carl Adams, Thomas Adams, Paolo Addesso, RX~Adhikari, et~al.
\newblock Observation of gravitational waves from a binary black hole merger.
\newblock {\em Physical review letters}, 116(6):061102, 2016.

\bibitem{abbott2019gwtc1}
BP~Abbott, R~Abbott, TD~Abbott, S~Abraham, F~Acernese, K~Ackley, C~Adams,
  RX~Adhikari, VB~Adya, C~Affeldt, et~al.
\newblock Gwtc-1: a gravitational-wave transient catalog of compact binary
  mergers observed by ligo and virgo during the first and second observing
  runs.
\newblock {\em Physical Review X}, 9(3):031040, 2019.

\bibitem{abbott2020gwtc2}
R~Abbott, TD~Abbott, S~Abraham, F~Acernese, K~Ackley, A~Adams, C~Adams,
  RX~Adhikari, VB~Adya, C~Affeldt, et~al.
\newblock Gwtc-2: Compact binary coalescences observed by ligo and virgo during
  the first half of the third observing run.
\newblock {\em arXiv preprint arXiv:2010.14527}, 2020.

\bibitem{abernathy2011einstein}
Matt Abernathy, F~Acernese, P~Ajith, B~Allen, P~Amaro~Seoane, N~Andersson,
  S~Aoudia, P~Astone, B~Krishnan, L~Barack, et~al.
\newblock Einstein gravitational wave telescope conceptual design study.
\newblock 2011.

\bibitem{ET2020einstein}
ET~Steering Committee~Editorial Team.
\newblock Einstein telescope design report update 2020.
\newblock 2020.

\bibitem{adhikari2020cryogenic}
R~X Adhikari and et. al.
\newblock A cryogenic silicon interferometer for gravitational-wave detection.
\newblock {\em Classical and Quantum Gravity}, 37(16):165003, jul 2020.

\bibitem{reitze2019cosmic}
David Reitze, Rana~X Adhikari, Stefan Ballmer, Barry Barish, Lisa Barsotti,
  GariLynn Billingsley, Duncan~A Brown, Yanbei Chen, Dennis Coyne, Robert
  Eisenstein, et~al.
\newblock Cosmic explorer: the us contribution to gravitational-wave astronomy
  beyond ligo.
\newblock {\em arXiv preprint arXiv:1907.04833}, 2019.

\bibitem{aasi2015advanced}
Junaid Aasi, BP~Abbott, Richard Abbott, Thomas Abbott, MR~Abernathy, Kendall
  Ackley, Carl Adams, Thomas Adams, Paolo Addesso, RX~Adhikari, et~al.
\newblock Advanced ligo.
\newblock {\em Classical and quantum gravity}, 32(7):074001, 2015.

\bibitem{acernese2014advanced}
Fet~al Acernese, M~Agathos, K~Agatsuma, D~Aisa, N~Allemandou, A~Allocca,
  J~Amarni, P~Astone, G~Balestri, G~Ballardin, et~al.
\newblock Advanced virgo: a second-generation interferometric gravitational
  wave detector.
\newblock {\em Classical and Quantum Gravity}, 32(2):024001, 2014.

\bibitem{somiya2012detector}
Kentaro Somiya.
\newblock Detector configuration of kagra -- the japanese cryogenic
  gra\-vitational-wave detector.
\newblock {\em Classical and Quantum Gravity}, 29(12):124007, 2012.

\bibitem{akutsu2020overview}
T~Akutsu, M~Ando, K~Arai, Y~Arai, S~Araki, A~Araya, N~Aritomi, Y~Aso, S-W Bae,
  Y-B Bae, et~al.
\newblock Overview of kagra: Detector design and construction history.
\newblock {\em arXiv preprint arXiv:2005.05574}, 2020.

\bibitem{buikema2020sensitivity}
A~Buikema, C~Cahillane, GL~Mansell, CD~Blair, R~Abbott, C~Adams, RX~Adhikari,
  A~Ananyeva, S~Appert, K~Arai, et~al.
\newblock Sensitivity and performance of the advanced ligo detectors in the
  third observing run.
\newblock {\em Physical Review D}, 102(6):062003, 2020.

\bibitem{acernese2020advanced}
F~Acernese, T~Adams, K~Agatsuma, L~Aiello, A~Allocca, A~Amato, S~Antier,
  N~Arnaud, S~Ascenzi, P~Astone, et~al.
\newblock Advanced virgo status.
\newblock In {\em Journal of Physics: Conference Series}, volume 1342, page
  012010. IOP Publishing, 2020.

\bibitem{akutsu2019first}
T~Akutsu, M~Ando, K~Arai, Y~Arai, S~Araki, A~Araya, N~Aritomi, H~Asada, Y~Aso,
  S~Atsuta, et~al.
\newblock First cryogenic test operation of underground km-scale
  gravitational-wave observatory kagra.
\newblock {\em Classical and Quantum Gravity}, 36(16):165008, 2019.

\bibitem{ushiba2021cryogenic}
Takafumi Ushiba, Tomotada Akutsu, Sakae Araki, Rishabh Bajpai, Dan Chen, Kieran
  Craig, Yutaro Enomoto, Ayako Hagiwara, Sadakazu Haino, Yuki Inoue, et~al.
\newblock Cryogenic suspension design for a kilometer-scale gravitational-wave
  detector.
\newblock {\em Classical and Quantum Gravity}, 38(8):085013, 2021.

\bibitem{shapiro2017cryogenically}
Brett Shapiro, Rana~X Adhikari, Odylio Aguiar, Edgard Bonilla, Danyang Fan,
  Litawn Gan, Ian Gomez, Sanditi Khandelwal, Brian Lantz, Tim MacDonald, et~al.
\newblock Cryogenically cooled ultra low vibration silicon mirrors for
  gravitational wave observatories.
\newblock {\em Cryogenics}, 81:83--92, 2017.

\bibitem{bonilla2019improving}
Edgard Bonilla and Brian Lantz.
\newblock Improving the cool-down times for third ge\-nera\-tion
  gra\-vita\-tional wave observatories (lvc).
\newblock \url{https://dcc.ligo.org/LIGO-G1900526/public}, 2019.
\newblock LIGO document, LIGO-G1900526-v1.

\bibitem{tobar1995university}
ME~Tobar, DG~Blair, EN~Ivanov, Frank Van~Kann, NP~Linthorne, PJ~Turner, and
  IS~Heng.
\newblock The university of western australia? s resonant-bar gravitational
  wave experiment.
\newblock {\em Australian Journal of Physics}, 48(6):1007--1026, 1995.

\bibitem{corruccini1959gaseous}
RJ~Corruccini.
\newblock Gaseous heat conduction at low pressures and temperatures.
\newblock {\em Vacuum}, 7:19--29, 1959.

\bibitem{cavalleri2010gas}
A~Cavalleri, G~Ciani, R~Dolesi, M~Hueller, D~Nicolodi, D~Tombolato, S~Vitale,
  PJ~Wass, and WJ~Weber.
\newblock Gas damping force noise on a macroscopic test body in an infinite gas
  reservoir.
\newblock {\em Physics Letters A}, 374(34):3365--3369, 2010.

\bibitem{trott2011experimental}
Wayne~M Trott, Jaime~N Casta{\~n}eda, John~R Torczynski, Michael~A Gallis, and
  Daniel~J Rader.
\newblock An experimental assembly for precise measurement of thermal
  accommodation coefficients.
\newblock {\em Review of scientific instruments}, 82(3):035120, 2011.

\bibitem{agrawal2008survey}
Amit Agrawal and SV~Prabhu.
\newblock Survey on measurement of tangential momentum accommodation
  coefficient.
\newblock {\em Journal of Vacuum Science \& Technology A: Vacuum, Surfaces, and
  Films}, 26(4):634--645, 2008.

\bibitem{cao2005temperature}
Bing-Yang Cao, Min Chen, and Zeng-Yuan Guo.
\newblock Temperature dependence of the tangential momentum accommodation
  coefficient for gases.
\newblock {\em Applied Physics Letters}, 86(9):091905, 2005.

\bibitem{naticchioni2018payloads}
L~Naticchioni, Virgo Collaboration, et~al.
\newblock The payloads of advanced virgo: current status and upgrades.
\newblock In {\em Journal of Physics: Conference Series}, volume 957, page
  012002. IOP Publishing, 2018.

\bibitem{gil2009cryogenic}
Woosik Gil, Jochen Bonn, Beate Bornschein, Rainer Gehring, Oleg Kazachenko,
  Jonny Kleinfeller, et~al.
\newblock The cryogenic pumping section of the katrin experiment.
\newblock {\em IEEE transactions on applied superconductivity}, 20(3):316--319,
  2009.

\bibitem{zucker1996measurement}
Michael~E Zucker and Stanley~E Whitcomb.
\newblock Measurement of optical path fluctuations due to residual gas in the
  ligo 40 meter interferometer.
\newblock In {\em Proceedings of the Seventh Marcel Grossman Meeting on recent
  developments in theoretical and experimental general relativity, gravitation,
  and relativistic field theories}, pages 1434--1436, 1996.

\bibitem{celestini2008cosine}
Franck Celestini and Fabrice Mortessagne.
\newblock Cosine law at the atomic scale: toward realistic simulations of
  knudsen diffusion.
\newblock {\em Physical Review E}, 77(2):021202, 2008.

\bibitem{gombosi1994gaskinetic}
Tamas~I Gombosi and Atmo Gombosi.
\newblock {\em Gaskinetic theory}, chapter 7.2.3.
\newblock Number~9. Cambridge University Press, 1994.

\bibitem{jeans_1982}
James Jeans.
\newblock {\em An Introduction to the Kinetic Theory of Gases}.
\newblock Cambridge Science Classics. Cambridge University Press, 1982.

\bibitem{schlamminger2010comparison}
Stephan Schlamminger.
\newblock Comparison of squeeze film damping simulations for the advanced ligo
  geometry.
\newblock \url{http://www.ligo.caltech.edu/docs}, 2010.
\newblock LIGO document, LIGO-T1000101-v1.

\bibitem{cavalleri2009increased}
A~Cavalleri, G~Ciani, R~Dolesi, A~Heptonstall, M~Hueller, D~Nicolodi, S~Rowan,
  D~Tombolato, S~Vitale, PJ~Wass, et~al.
\newblock Increased brownian force noise from molecular impacts in a
  constrained volume.
\newblock {\em Physical review letters}, 103(14):140601, 2009.

\bibitem{weiss2009gas}
Rainer Weiss.
\newblock Gas damping of the final stage in the advanced ligo suspensions.
\newblock \url{https://dcc.ligo.org/LIGO-T0900509/public}, 2009.
\newblock LIGO document, LIGO-T0900509-v1.

\bibitem{saulson1990thermal}
Peter~R Saulson.
\newblock Thermal noise in mechanical experiments.
\newblock {\em Physical Review D}, 42(8):2437, 1990.

\bibitem{evans2011gas}
Matthew Evans, Peter Fritschel, Rai Weiss, and LIGO~Scientific Collaboration.
\newblock Gas damping monte carlo.
\newblock Technical report, Tech. Rep. LIGO-T0900582, 2011.

\end{thebibliography}

\end{document}